\documentclass[letter, 11pt, psamsfonts]{amsart}

\usepackage{titletoc}
\usepackage{amsmath,amsfonts,amssymb}
\usepackage{graphicx,wrapfig}
\usepackage{hyperref}
\usepackage{setspace}

\title[interaction between baroreflex and cerebral autoregulation]{Mathematical model of the interaction between baroreflex and cerebral autoregulation}

\author[A Mahdi, MS Olufsen, SJ Payne]{Adam Mahdi$^1$, Mette S. Olufsen$^2$, Stephen J. Payne$^1$}

\begin{document}

\maketitle

\vspace{-0.5cm}

{\small
\begin{center}
$^1$ Institute of Biomedical Engineering, University of Oxford, UK\\
adam.mahdi@eng.ox.ac.uk,\quad stephen.payne@eng.ox.ac.uk
\end{center}

\begin{center}
$^2$ Department of Mathematics, NC State University, Raleigh 26795, USA\\
msolufse@ncsu.edu
\end{center}
}

\onehalfspacing

\begin{abstract}
Baroreflex (BR) and cerebral autoregulation (CA) are two important mechanisms regulating blood pressure and flow.  However, the functional relationship between BR and CA in humans is unknown. Since BR impairment is an adverse prognostic indicator for both cardiac and cerebrovascular diseases it would be of clinical interest to better understand the relationship between BR and CA. Motivated by this observation we develop a simple mathematical framework aiming to simulate the effects of BR on the cerebral blood flow dynamics.
\end{abstract}

\section{Introduction}
 \emph{Baroreflex} (BR) is  the main short-term blood pressure (BP) regulation mechanism of the cardiovascular  system (CVS). It aims to provide adequate perfusion of all tissues by maintaining blood flow and pressure at homeostasis by regulating heart rate (HR), vascular resistance, compliance and other variables of the  CVS.  \emph{Cerebral autoregulation} (CA) is a physiological mechanism which aims to maintain blood flow in the brain at an appropriate level  during changes in  BP. It is achieved by regulating cerebral arteriolar vessels to match the cerebral blood flow (CBF) with the metabolic demands of the brain. 

Although it has been known that both BR and CA are central in maintaining appropriate CBF the functional relationship between the two mechanisms in humans is unknown. Discerning the fundamental links between BR function and CBF regulation is of clinical significance, because BR impairment is a negative prognostic factor for cardiac and cerebrovascular disease \cite{Tzeng2010, Robinson2003}.  Motivated by this we develop a mathematical framework aiming to simulate the effects of BR on the cerebral blood flow dynamics. The novelty of the approach lies in inferring the neural activities (sympathetic and parasympathetic)  from the HR model and integrating them with the haemodynamic model and the CBF regulation. The framework is based on the current understanding of the physiology underlying the regulation mechanisms (both BR and CA) and allows to hypothesize the relative influence of sympathetic and parasympathetic firing rate on the CBF dynamics.


\section{Methodology}

\subsection{Heart rate regulation}
The control of HR is an important aspect of BR for maintaining homeostasis. Typically, BR is divided into an afferent part, a control center, and an efferent part \cite{Levick10}. The firing rate of the afferent baroreceptor neurons is modulated by changes in the viscoelastic stretch of the nerve endings which are embedded in the arterial wall of the aorta and carotid sinuses. The sympathetic firing rate increases the concentration of noradrenaline (NAd) via  a slow pathway which, in turn, raises the HR. Parasympathetic firing rate, on the other hand, increases the concentration of acetylcholine (ACh) which is released through a slow and fast pathways that form the total concentration leading to a reduction of HR.

\begin{figure}[t]
\begin{center}
\includegraphics[width=15.5cm]{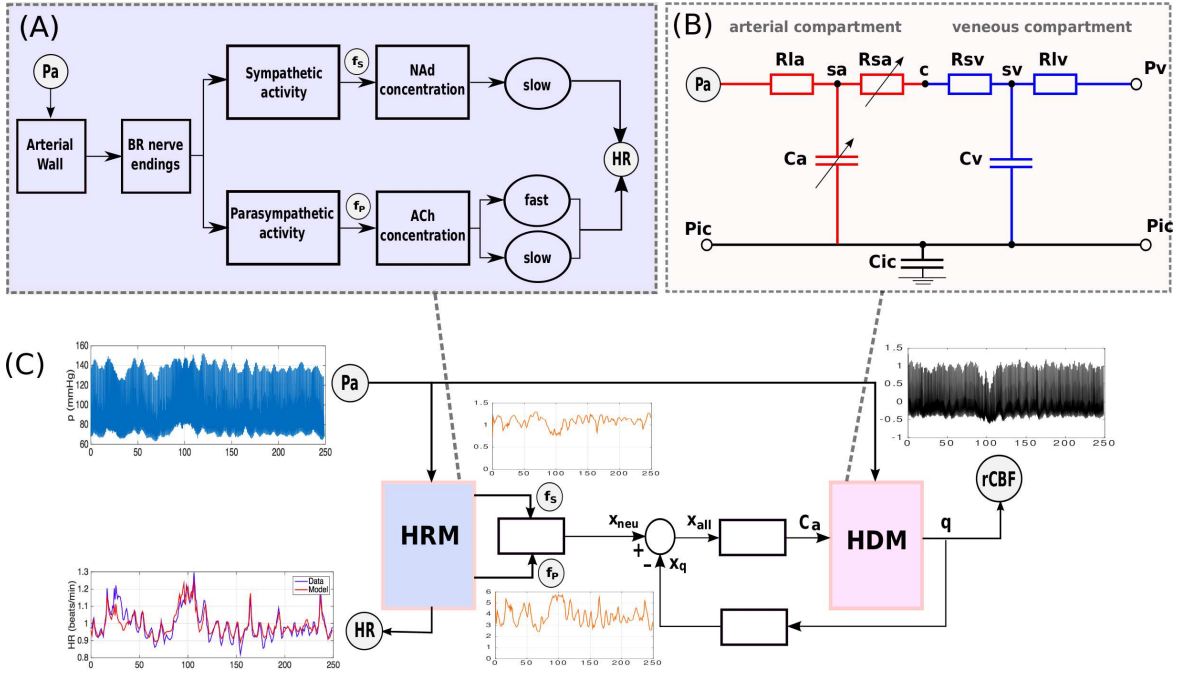}\hspace{1cm}
\caption{{\bf Model of interaction between BR and CA.} {(A):} {\it Baroreflex regulation of heart rate model (HRM).} Arterial blood pressure $P_a$ causes the changes in arterial wall strain and triggers a baroreceptor firing rate. The sympathetic $f_S$ and parasympathetic $f_P$ activities are generated within NTS and are involved in the heart rate control by modulating the release of acetylcholine, from the vagal nerves, and noradrenaline from the postganglionic sympathetic nerves. 
{\it (B):} {\it Haemodynamic model (HDM).} Electrical analogue of the cerebral haemodynamic and intracranial dynamics. {(C)} {\it  CBF  regulation mechanisms.} The schematic representation of the assumed interaction between HRM and HDM.}\label{mod:HRM}
\end{center}
\end{figure}

Figure~\ref{mod:HRM}(A) shows the block structure of the BR regulation of HR model (denoted HRM) used here to infer the sympathetic $f_S$ and parasympathetic $f_P$ activities by fitting the HR prediction of HRM to human HR data. For more details  and the basal values of the parameters see \cite{Ottesen2014, MahStuOttOlu13}.

\subsection{Haemodynamic model}
Figure~\ref{mod:HRM}(B)  depicts electrical circuits representation of the haemodynamic model (denoted HDM) used in this work. It includes the arterial, capillary and venous compartments. This division reflect some fundamental functions played by each compartment such as regulation of CBF (arteries), gas transport (capillaries), regulation of CBV  (veins).  The model assumes constant intracranial pressure, which is a reasonable assumption due to a relatively short time-series data used in our simulation. The autoregulation mechanism is achieved by  changing the compliance $C_a$ (discussed later in more detail) which, in turn, changes the resistance of the small arteries $R_{sa}$.  Since that the objective of the present work is to show qualitatively the influence of BR on CA, the computation of CBF via HDM can be thought of as performed in arbitrary units. This relative CBF (denoted rCBF) is computed as
\begin{equation}\label{rCBF}
q =\frac{P_{sa} - P_{sv}}{R_{sa} + R_{sv}},
\end{equation}
where
$P_{sa}$ and $P_{sv}$ is the cerebral blood pressure at the nodes $(sa)$ and $(sv)$, see Figure~\ref{mod:HRM}(B). The venous $C_v$ and intracranial $C_{ic}$ compliances as well as large arterial $R_{la}$ and large venous $R_{lv}$ resistances are assumed constant.  For  the basal values of the parameters and other details see similar approaches employed previously in, e.g., \cite{UL97, Payne06}.

\subsection{Feedback mechanisms}
The flow of blood to the brain is modulated both by local and global mechanisms including neurovascular coupling, changes in arterial partial pressures of oxygen and carbon dioxide, and the CBF feedback mechanisms. These  regulatory responses act collectively to maintain CBF and oxygen supply to the brain amidst changes in arterial BP.

 Figure~\ref{mod:HRM}(C) shows the conceptual model of the interaction between HRM and HDM, indicating the CBF regulation mechanisms implemented in our simulations. As a first step, we assume here that the changes in partial pressure of oxygen and carbon dioxide are constant.
The CBF feedback regulation is assumed to act with a delay via first order filter, with  a time-constant $\tau_q$, of the form
\begin{equation}\label{Reg:CBF}
\tau_q \frac{dx_q}{dt}=-x_q + G_q (q-q^o)/q^o,
\end{equation}
where $q^o$ is the baseline value of the CBF and $G_q$ is the adjustable CBF feedback gain.  The reduction of this gain can be interpreted as  the impairment of the CBF regulation.

Following \cite{Behzadi2005, Payne06}  the neural-based CBF regulation mechanism is modeled as the second-order differential equation
\begin{equation}\label{Reg:CO2}
\tau_{n1} \frac{d^2x_{neu}}{dt^2} + \tau_{n2} \frac{dx_{neu}}{dt}+x_{neu}= G_{neu} u(f_S,f_P),
\end{equation}
where $\tau_{n1}$ and $\tau_{n2}$ are the two time-constants, and $G_{neu}$ is the gain of the neural-based stimulation. The function $u(f_S,f_P)$ combines the influence of sympathetic $f_S$ and parasympathetic $f_P$ activities and is modeled as the following weighted sum 
\begin{equation}
u(f_S,f_P) = \alpha f_S + \beta f_P,
\end{equation}
where $\alpha$ and $\beta$ are the normalized proportionality constants such that $|\alpha|+|\beta|=1$.  This produces a quantity $x_{neu}$, which represents the neural-based contribution to the CBF regulation mechanism.
The changes in arterial compliance will be influenced by a combination of the above regulatory responses, i.e. the CBF feedback $x_q$ and the neural-based $x_{neu}$. We consider here the simplest case by assuming that all the mechanisms act in a linearly additive way 
\[
x_{all} = x_{neu}-x_q.
\]
Following \cite{Payne06, UL97} the influence of all the contributions of the regulatory responses $x_{all}$ on the arterial compliance $C_a$ is modeled using an asymmetrical sigmoidal function with upper and lower saturation levels: 
{\small
\[
C_a =  C^o_a +
\left\{\begin{array}{cc}
\Delta C_a^+ \tanh\Big(\dfrac{2 x_{all}}{\Delta C_a^+}\Big),&x_{all}\geq 0\\
\\
\Delta C_a^- \tanh\Big(\dfrac{2 x_{all}}{\Delta C_a^-}\Big),&x_{all}< 0,
\end{array}\right.
\]}
where $C^o_a$ is the baseline arterial compliance; and  $\Delta C_a^+$ (resp. $\Delta C_a^-$) is the amplitude of positive (resp. negative) change in arterial compliance.

\section{Results and conclusions}

\begin{figure}[t!]
\begin{center}
\includegraphics[width=7.5cm]{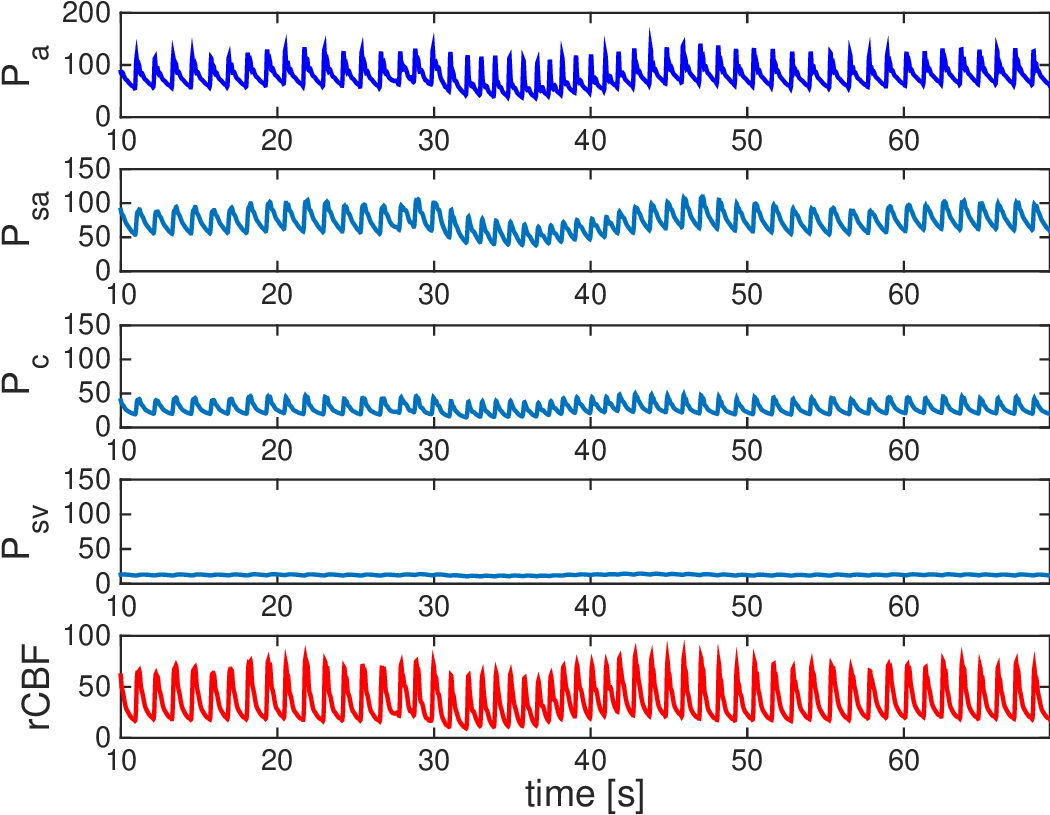}\quad
\includegraphics[width=7.5cm]{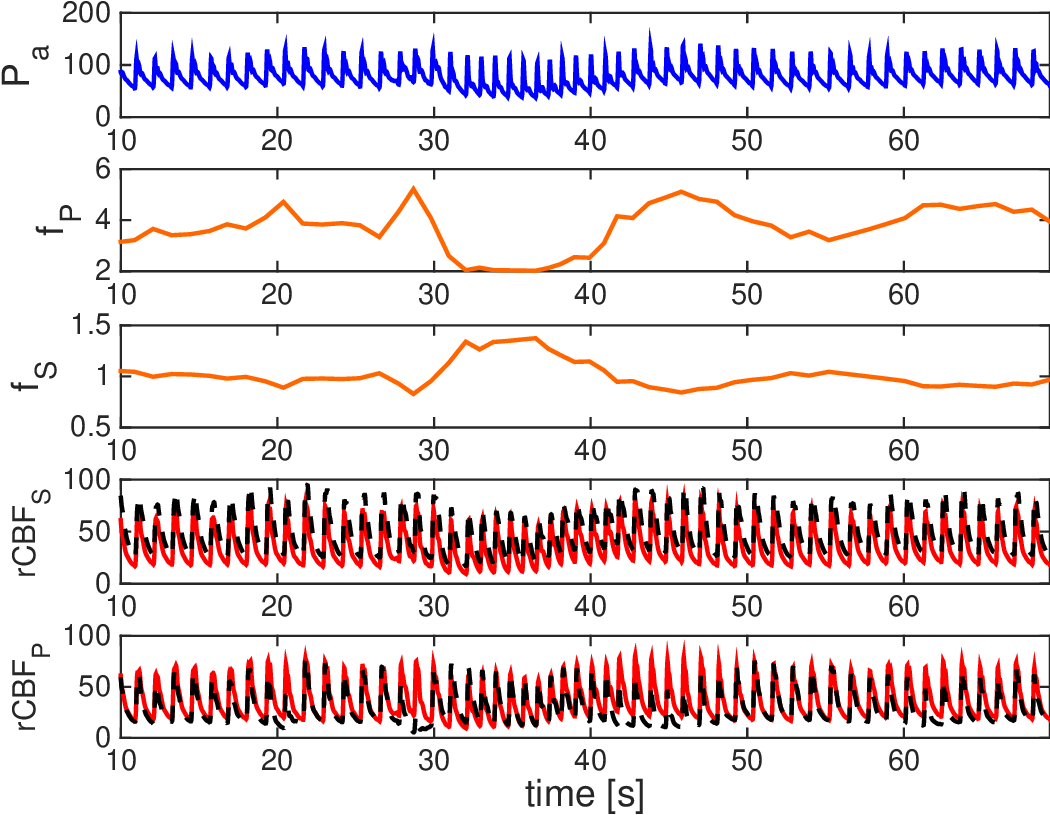}
\caption{{\bf Neural-based simulation during sit-to-stand experiment.} {\it Left:} Blood pressure simulations at various nodes of  the haemodynamic model (HDM) and the corresponding relative cerebral blood flow (rCBF) changes  during sit-to-stand experiment with no neural activities, i.e. $G_{neu}=0$. Here $P_a$  is the input pressure (data) and  $P_j$  is the HDM-based simulated pressure at the node  $j\in\{sa, c, sv\}$, see Figure~\ref{mod:HRM}(B). \qquad\qquad
{\it Right:}  The input pressure $P_a$ and the simulated sympathetic $f_S$ and parasympathetic $f_P$ activity estimated by fitting the heart rate model (HRM), see in Figure~\ref{mod:HRM}(A), to heart rate (data). Finally, the simulated rCBF under dominant sympathetic ($f_S>f_P$) and dominant parasympathetic ($f_P>f_S$) activity are denoted by $rCBF_S$  and $rCBF_P$ (dotted black), respectively; while continuous red line shows rCBF ($G_{neu}=0$) and is plotted to facilitate the comparison.}\label{res:rCBF} 
\end{center}
\end{figure}

Figure~\ref{res:rCBF}(left) shows the simulation of the BP at various nodes of the haemodynamic model HDM (see Figure~\ref{mod:HRM}(B)) and the corresponding rCBF (see \eqref{rCBF}) during sit-to-stand experiment assuming no neural activities, i.e. $G_{neu}=0$. Here $P_a$  is the input pressure (data) and  $P_j$, where  $j\in\{sa, c, sv\}$  is the simulated pressure at various nodes indicated in  Figure~\ref{mod:HRM}(B).

Figure~\ref{res:rCBF}(right) shows the input pressure ($P_a$) and the inferred sympathetic $f_S$ and  parasympathetic $f_P$ activity, which are estimated by fitting HRM, schematically represented in Figure~\ref{mod:HRM}(A), to the BP/HR human data. The simulated rCBF under dominant sympathetic ($f_S>f_P$) and parasympathetic ($f_P>f_S$) activity are denoted by $rCBF_S$  and $rCBF_P$ (dotted black), respectively; these are plotted agains rCBF with no neural activity, i.e. $G_{neu}=0$ (continuous red), to facilitate the comparison of the simulated relative changes of the CBF. 

{\it Acknowledgement.} A.M. and S.J.P. acknowledge the support of the EPSRC project EP/K036157/1 and M.S.O. acknowledge the support of the VPR project under NIH-NIGMS grant 1P50-GM094503-01A0 subaward to NCSU. The authors thank Dr. L. Lipsitz at the Hebrew SeniorLife in Boston, MA. for providing  anonymized patient  data.

\end{document}